\documentclass[onecolumn]{IEEEtran}
\usepackage{amsmath}
\usepackage{amsfonts}
\usepackage{amssymb}
\usepackage{algorithm}
\usepackage{algorithmic}
\usepackage{graphicx}
\usepackage{verbatim}

\newtheorem{theorem}{Theorem}
\newtheorem{lemma}{Lemma}
\newtheorem{remark}{Remark}

\newtheorem{corollary}{Corollary}
\newtheorem{example}{Example}

\newcommand{\beq}{\begin{equation}}
\newcommand{\eeq}{\end{equation}}
\newcommand{\beqnn}{\begin{equation*}}
\newcommand{\eeqnn}{\end{equation*}}
\newcommand{\beqy}{\begin{eqnarray}}
\newcommand{\eeqy}{\end{eqnarray}}
\newcommand{\beqynn}{\begin{eqnarray*}}
\newcommand{\eeqynn}{\end{eqnarray*}}
\newcommand{\bit}{\begin{itemize}}
\newcommand{\eit}{\end{itemize}}
\newcommand{\ben}{\begin{enumerate}}
\newcommand{\een}{\end{enumerate}}
\newcommand{\bex}{\begin{example}}
\newcommand{\eex}{\end{example}}

\newcommand{\balg}[1]{\begin{algorithm} \caption{#1}}
\newcommand{\ealg}{\end{algorithm}}

\newcommand{\balgc}{\begin{algorithmic}[1]}
\newcommand{\ealgc}{\end{algorithmic}}

\newcommand{\bary}{\begin{array}}
\newcommand{\eary}{\end{array}}
\newcommand{\bmx}{\begin{bmatrix}}
\newcommand{\emx}{\end{bmatrix}}
\newcommand{\bsmx}{\left[\begin{smallmatrix}}
\newcommand{\esmx}{\end{smallmatrix}\right]}
\newcommand{\bmxc}[1]{\left[\begin{array}{@{}#1@{}}}
\newcommand{\emxc}{\end{array}\right]}
\newcommand{\bcn}{\begin{center}}
\newcommand{\ecn}{\end{center}}

\newcommand{\A}{\boldsymbol{A}}

\newcommand{\e}{\boldsymbol{e}}

\newcommand{\h}{\boldsymbol{h}}

\renewcommand{\u}{\boldsymbol{u}}

\newcommand{\x}{{\boldsymbol{x}}}
\newcommand{\y}{{\boldsymbol{y}}}

\usepackage{slashbox}

\begin{document}
%
\title{Stable Recovery of Sparse Signals via $l_p-$Minimization}

\author{Jinming~Wen, Dongfang~Li and~Fumin~Zhu
\thanks{Jinming~Wen is with The Department of Mathematics and Statistics,
McGill University, Montreal, QC H3A 0B9, Canada (e-mail: jinming.wen@math.mcgill.ca); Dongfang~Li is with the School of Mathematics and Statistics, Huazhong University of Science and Technology, Wuhan, China, 430074 (e-mail: dfli@hust.edu.cn); Fumin~Zhu is with the School of Economic Information Engineering, Southwestern University of Finance and Economics, he is also with the Department of Applied Mathematics and Statistics, State University of New York at Stony Brook, NY 11794, Chengdu, China, 611130 (e-mail: zhufumin520@163.com).}

\thanks{Manuscript received; revised .}}



\maketitle

\begin{abstract}

In this paper, we show that, under the assumption that $\|\e\|_2\leq \epsilon$, every $k-$sparse signal $\x\in \mathbb{R}^n$ can be stably ($\epsilon\neq0$) or exactly recovered ($\epsilon=0$) from $\y=\A\x+\e$ via $l_p-$mnimization with $p\in(0, \bar{p}]$, where
\beqnn
\bar{p}=
\begin{cases}
\frac{50}{31}(1-\delta_{2k}), &\delta_{2k}\in[\frac{\sqrt{2}}{2}, 0.7183)\cr
0.4541, &\delta_{2k}\in[0.7183,0.7729)\cr
2(1-\delta_{2k}), &\delta_{2k}\in[0.7729,1)
\end{cases},
\eeqnn
even if the restricted isometry constant of $\A$ satisfies $\delta_{2k}\in[\frac{\sqrt{2}}{2}, 1)$.
Furthermore, under the assumption that $n\leq 4k$, we show that the range of $p$ can be further improved to $p\in(0,\frac{3+2\sqrt{2}}{2}(1-\delta_{2k})]$. This not only extends some discussions of only the noiseless recovery (Lai et al. and Wu et al.) to the noise recovery, but also greatly improves the best existing results where $p\in(0,\min\{1, 1.0873(1-\delta_{2k}) \})$ (Wu et al.).
\end{abstract}


\section{Introduction}

In compressed sensing, see, e.g., \cite{CanT05}, \cite{Don06}, \cite{CohDD09}, the following linear model is observed:
\beq
\label{e:model}
\y=\A\x+\e
\eeq
where $\x\in \mathbb{R}^n$ is an unknown signal, $\y\in \mathbb{R}^m$ is an observation vector, $\A\in \mathbb{R}^{m\times n}$ (with $m<<n$) is a known sensing matrix and $\e \in \mathbb{R}^{m}$ is the measurement error vector. For simplicity, in this paper, we only consider $l_2$ bounded noise, i.e., $\|\e\|_2\leq \epsilon$ for some $\epsilon$, see, e.g., \cite{Fuc05}, \cite{DonET06}, \cite{Can08}. If there is no noise, we take $\epsilon=0$.


One of the central goals of compressed sensing is to recover $\x$ based on $\A$ and $\y$. It has been shown that under some suitable conditions, $\x$ can be stably or exactly recovered, see, e.g., \cite{CanRT06},  \cite{Mol11}.

A common method to recover $\x$ from \eqref{e:model} is to solve the following $l_1-$minimization problem:
\begin{eqnarray}
\label{e:l1}
\min_{\gamma\in \mathbb{R}^n}\|\gamma\|_1 :\; \;\text{subject \;to} \;\|\y-\A\gamma\|_2\leq \epsilon.
\end{eqnarray}

One of the commonly used frameworks for sparse recovery is the restricted isometry property (RIP)
which was introduced in \cite{CanT05}. A vector $\x\in \mathbb{R}^n$ is $k-$sparse if $|\text{supp}(\x)|\leq k$, where
$\text{supp}(\x)=\{i:x_i\neq0\}$ is the support of $\x$.
For any $m\times n$ matrix $\A$ and any integer $k, 1\leq k\leq n$, the $k-$restricted isometry constant (RIC) $\delta_k$ is defined as the smallest
constant such that
\beq
\label{e:RIP}
(1-\delta_k)\|\x\|_2^2\leq \|\A\x\|_2^2\leq(1+\delta_k)\|\x\|_2^2
\eeq
for all $k-$sparse vector $\x$.
If $k+k'\leq n$, then the $k,k'-$restricted orthogonality constant (ROC) $\theta_{k,k'}$ is defined as the smallest constant such that
\beqnn
|\langle\A\x, \A\x'\rangle|\leq \theta_{k,k'}\|\x\|_2\|\x'\|_2
\eeqnn
for all $\x$ and $\x'$, where $\x$ and $\x'$  are respectively $k-$sparse and $k'-$sparse and have disjoint supports.


A variety sufficient conditions based on RIC and ROC for the stable recovery ($\epsilon\neq0$) or exact recovery ($\epsilon=0$) of $k-$sparse signal $\x$
have been introduced in the literature. For example, $\delta_k+\theta_{k,k}+\theta_{k,2k}<1$ in \cite{CanT05} and $\delta_{2k}+\theta_{k,2k}<1$ in \cite{CanT07}.
Sufficient conditions based on only RIC have also been given. For example, $\delta_{3k}+3\delta_{4k}<2$ and $\delta_{k}<\frac{1}{3}$ were respectively given in \cite{CanRT06} and \cite{CaiZ13}.
The sufficient conditions also include $\delta_{2k}<\sqrt{2}-1$ in
\cite{Can08}, $\delta_{2k}<0.4531$ in \cite{FouL09}, $\delta_{2k}<0.4652$ in \cite{Fou10b} and $\delta_{2k}<\frac{\sqrt{2}}{2}$ in \cite{CaiZ13b}.

However, it was shown in \cite{DavG09} that exactly recover $\x$ is not always possible if $\delta_{2k}\geq\frac{\sqrt{2}}{2}$. Therefore, one chooses to solve \eqref{e:lp} with $p\in (0,1)$ to recover $\x$, see, e.g., \cite{Sun12} for $\epsilon\neq0$ and
\cite{FouL09}, \cite{LaiL11}, \cite{Wuc13} for $\epsilon=0$.
\begin{eqnarray}
\label{e:lp}
\min_{\gamma\in \mathbb{R}^n}\|\gamma\|_p :\; \;\text{subject \;to} \;\|\y-\A\gamma\|_2\leq \epsilon.
\end{eqnarray}

Although the $l_p-$minimization problem is more difficult to solve than the $l_1-$minimization problem  due to its non-convexity and non-smoothness \cite{Sun12}, there are some efficient algorithms to solve \eqref{e:lp}, see, e.g., \cite{FouL09}, \cite{DauDFG10} and \cite{Sun12}.

The $l_p-$minimization requires weaker condition on $\delta_{2k}$ than that of the $l_1-$minimization.
It was shown in \cite{Cha07} that for any $\delta_{2k+1}\in (0,1)$, there is some $p$ such that one can exactly recover the $k-$sparse signal $\x$ via solving \eqref{e:lp} with $\epsilon=0$.
In \cite{Sun12}, Sun showed that for any $\delta_{2k}\in (0,1)$, one can stably recover ($\epsilon\neq0$) or exactly recover ($\epsilon=0$) the $k-$sparse signal $\x$ via solving \eqref{e:lp}, where $p$ is about $0.6797(1-\delta_{2k})$. For the noiseless recovery, the range of $p$ has been improved to $p<\min\{1, 1.0873(1-\delta_{2k}) \}$ in \cite{Wuc13}.


As far as we know, $p<\min\{1, 1.0873(1-\delta_{2k}) \}$ is the best existing results. Therefore, a natural question is to ask whether this condition can be further improved. If so, can the improved condition be extended to the noise recovery?

The answers are affirmative. If $\delta_{2k}<\frac{\sqrt{2}}{2}$, then one can choose $p=1$ \cite{CaiZ13b}. Therefore, we only need to improve the range of $p$ for each given $\delta_{2k}\in[\frac{\sqrt{2}}{2}, 1)$. In this paper, we will show that for each given
$\delta_{2k}\in[\frac{\sqrt{2}}{2},1)$ for general $k$, one can stably recover ($\epsilon\neq0$) or exactly recover ($\epsilon=0$) the $k-$sparse signal $\x\in \mathbb{R}^n$ via solving \eqref{e:lp} with $p\in(0, \bar{p}]$, where $\bar{p}$ is defined in \eqref{e:pbar}.
Under the assumption that $k\geq\frac{n}{4}$, we will show that the range of $p$ can be further improved to $p\in(0,\frac{3+2\sqrt{2}}{2}(1-\delta_{2k})]$. This will not only extend some discussions of only the noiseless recovery \cite{LaiL11}, \cite{Wuc13} to the noise recovery, but will also greatly improve the best existing results where $p<\min\{1, 1.0873(1-\delta_{2k}) \}$ \cite{Wuc13}.

The rest of the paper is organized as follows.
In section II, we will give our main results.
In section III, we will develop some new techniques to prove the main results. Finally  we summarize this paper in section \ref{s:con}.

\section{Main Results}

\subsection{Preliminaries}

Suppose $\x$ in \eqref{e:model} is the real signal which we need to recover and $\x^{\star}$ is the solution of the $l_p$ minimization problem \eqref{e:lp}. Like in \cite{Can08}, we set $\h=\x-\x^{\star}$ and denote its $i-$th ($1\leq i\leq n$) component by $h_i$.
Similar to the notation used in \cite{Wuc13}, we respectively assume $T_0$ be the set $\{1,2,\ldots, k\}$, $T_0^c$ be the set $\{k+1,k+2,\ldots, n\}$ and $\x_{T_0^c}$ be the vector equal to $\x$ on the index set $T_0^c$ and zero elsewhere.
As assumed in \cite{Mol11} and \cite{Wuc13}, for simplicity, we assume that $\h_{T_0^c}$ is already sorted in non-increasing order of magnitude, i.e.,
$|h_{k+1}|\geq |h_{k+2}|\geq\ldots\geq |h_{n}|$. We also assume that $n=(l+1)k$ with $l$ being a positive integer. Partition the index set $T_0^c$ as the union of the subsets $T_i=\{ik+1,ik+2,\ldots, (i+1)k\}$ with $i\in\{1,2,\ldots, l\}$.


Let
\beq
\label{e:fp}
f(p)=(\frac{p}{2})^{\frac{1}{2}}(2-p)^{\frac{1}{p}-\frac{1}{2}}, p\in(0,1],
\eeq
\begin{align}
\label{e:gp}
g(p)=\frac{p}{2}(1-\frac{p}{2})^{\frac{2}{p}-1},p\in(0,1].
\end{align}
By some simple calculations, we have
\begin{eqnarray*}
(\ln f(p))'=-\frac{1}{p^2}\ln(2-p)\leq0, &(\ln g(p))'=-\frac{2}{p^2}\ln(1-\frac{p}{2})>0,\\
f(1)=\sqrt{2}/2, \lim_{p\rightarrow0^+}f(p)=+\infty, &g(1)=\frac{1}{4},\; \lim\limits_{p\rightarrow0^+}g(p)=0.
\end{eqnarray*}
Therefore,
\beq
\label{e:gpr}
g(p)\in (0,\frac{1}{4}), \;\forall p\in(0,1),
\eeq
and $f(p)=1$ has a unique solution. Let $p^{\star}$ be the unique solution of $f(p)=1$, then
\beq
\label{e:pstar}
p^{\star}\approx 0.45418,
\eeq
and
\begin{eqnarray}
\label{e:fpr}
\begin{cases}
f(p)\in[1,+\infty), &p\in(0,p^{\star}]\cr
f(p)\in[\sqrt{2}/2,1), &p\in(p^{\star},1]
\end{cases}.
\end{eqnarray}
By \eqref{e:fp} and the aforementioned equation, we have
\beq
\label{e:fpp}
p^{\frac{1}{2}}(2-p)^{\frac{1}{p}-\frac{1}{2}}>1, \;\text{if}\; p\in(p^{\star},1).
\eeq


In the following, we will give our main results. Like in \cite{Mol11}, we divide them into two cases: general case and special case ($n\leq4k$).

\subsection{General Case}

Let
\begin{align}
\label{e:Cp}
C(p)=
\begin{cases}
(\frac{((2-\delta_{2k})^{1-\frac{2}{p}}+2\delta_{2k})g(p)}{1-\delta_{2k}})^{p/2}, &p\in(0,p^{\star}]\cr
(\frac{(2-\delta_{2k})^{1-\frac{2}{p}}g(p)+2^{2-\frac{2}{p}}\delta_{2k}}{1-\delta_{2k}})^{p/2}, &p\in(p^{\star},1)
\end{cases},
\end{align}
where $g(p)$ is defined as in \eqref{e:gp}. Then we have the following result whose proof will be provided in Section \ref{s:12}.
\begin{theorem} \label{t:pnorm}
Suppose that $\A$ and $\e$ in \eqref{e:model} respectively satisfy the RIP with given $\delta_{2k}\in[\frac{\sqrt{2}}{2}, 1)$ and $\|\e\|_2\leq \epsilon$, then for each $p\in(0,1)$ such that
\begin{align}
\label{e:Cp1}
C(p)<1,
\end{align}
the solution $\x^{\star}$ to the $l_p-$minimization problem \eqref{e:lp} obeys
\begin{eqnarray}
\label{e:pnorm}
\|\x-\x^{\star}\|_p^p\leq C_0\|\x_{T_0^c}\|_p^p+C_1k^{1-\frac{p}{2}}\epsilon^p,
\end{eqnarray}
where
\begin{align*}
C_0=\frac{2(1+C(p))}{1-C(p)},\; C_1=\frac{2^{\frac{3p}{2}+1}}{(1-\delta_{2k})^{\frac{p}{2}}(1-C(p))}
\end{align*}
with $C(p)$ defined as in \eqref{e:Cp}. In particular, if $\epsilon=0$ and $\x$ is $k-$sparse, then the recovery is exact.
\end{theorem}

For a given $\delta_{2k}\in[\frac{\sqrt{2}}{2}, 1)$, it may be complicate to determine the range of $p$ such that \eqref{e:Cp1} holds, so in the
following, we would like to give a simple rule to determine it. To do this, we need to introduce the following lemma.

For each $p\in(0,1)$, let
\begin{align}
\label{e:hp}
h(p)=
\begin{cases}
-0.5p+1, &p\in(0,p^{\star}]\cr
-0.62p+1, &p\in(p^{\star},1)
\end{cases},
\end{align}
then we have the following result whose proof will be provided in Appendix \ref{s:lem1}.

\begin{lemma} \label{l:hp}
For each given $\delta_{2k}\in[\frac{\sqrt{2}}{2}, 1)$, if $\delta_{2k}\leq h(p)$, then \eqref{e:Cp1} holds.
\end{lemma}

By \eqref{e:pstar} and \eqref{e:hp}, we have
\beqnn
\min_{p\in(0,p^{\star}]} h(p)\geq 1-0.5p^{\star}>1-0.5\times0.4542=0.7729; \max_{p\in(p^{\star},1)} h(p)>1-0.62\times0.4542>0.7183.
\eeqnn
Therefore, for each given $\delta_{2k}\in[\frac{\sqrt{2}}{2}, 1)$, if $\delta_{2k}\in[\frac{\sqrt{2}}{2}, 0.7183)$, then $\delta_{2k}\leq h(p)$ holds for each $p\in(0,\frac{50}{31}(1-\delta_{2k})]$; if $\delta_{2k}\in[0.7183,0.7729)$, then $\delta_{2k}\leq h(p)$ holds for each $p\in(0,p^{\star}]$; and if $\delta_{2k}\in[0.7729,1)$, then $\delta_{2k}\leq h(p)$ holds for each $p\in(0,2(1-\delta_{2k})]$.

For each $\delta_{2k}\in[\frac{\sqrt{2}}{2}, 1)$, let
\begin{align}
\label{e:pbar}
\bar{p}=
\begin{cases}
\frac{50}{31}(1-\delta_{2k}), &\delta_{2k}\in[\frac{\sqrt{2}}{2}, 0.7183)\cr
p^{\star}, &\delta_{2k}\in[0.7183,0.7729)\cr
2(1-\delta_{2k}), &\delta_{2k}\in[0.7729,1)
\end{cases},
\end{align}
then by the aforementioned analysis, $\delta_{2k}\leq h(p)$ holds for each $p\in(0,\bar{p}]$. Therefore, by Theorem \ref{t:pnorm}
and Lemma \ref{l:hp}, we immediately have the following result.

\begin{corollary} \label{c:pnorm}
Suppose that $\A$ and $\e$ in \eqref{e:model} respectively satisfy the RIP with given $\delta_{2k}\in[\frac{\sqrt{2}}{2}, 1)$ and $\|\e\|_2\leq \epsilon$,
then for each $p\in(0,\bar{p}]$, where $\bar{p}$ is defined as in \eqref{e:pbar}, the solution $\x^{\star}$ to the $l_p-$minimization problem \eqref{e:lp} obeys \eqref{e:pnorm}. In particular, if $\epsilon=0$ and $\x$ is $k-$sparse, then the recovery is exact.
\end{corollary}

Theorem \ref{t:pnorm} and Corollary \ref{c:pnorm} give the bound on the $p-$norm of the error. Like in \cite{Can08}, we also
want to bound the $2-$norm of the error.
Let
\begin{eqnarray}
\label{e:Dp}
D(p)=
\begin{cases}
(\frac{(2+\delta_{2k})g(p)}{1-\delta_{2k}})^{p/2}, &p\in(0,p^{\star}]\cr
(\frac{(2-\delta_{2k})g(p)+2^{2-\frac{2}{p}}\delta_{2k}}{1-\delta_{2k}})^{p/2}, &p\in(p^{\star},1)
\end{cases},
\end{eqnarray}
where $g(p)$ is defined as in \eqref{e:gp}. Then we have the following result whose proof will be provided in Section \ref{s:12}.

\begin{theorem} \label{t:2norm}
Suppose that $\A$ and $\e$ in \eqref{e:model} respectively satisfy the RIP with given $\delta_{2k}\in[\frac{\sqrt{2}}{2}, 1)$ and $\|\e\|_2\leq \epsilon$, then for each $p\in(0,1)$ such that \eqref{e:Cp1} holds, the solution $\x^{\star}$ to the $l_p-$minimization problem \eqref{e:lp} obeys
\begin{eqnarray}
\label{e:2norm}
\|\x-\x^{\star}\|_2^p\leq D_0k^{\frac{p}{2}-1}\|\x_{T_0^c}\|_p^p+D_1\epsilon^p,
\end{eqnarray}
where
\begin{align*}
D_0=\frac{2D(p)}{1-C(p)},\,D_1=\frac{1}{(1-\delta_{2k})^{\frac{p}{2}}}(2^p+\frac{2^{\frac{3p}{2}}D(p)}{1-C(p)})
\end{align*}
with $C(p)$ and $D(p)$ defined as in \eqref{e:Cp} and \eqref{e:Dp}, respectively.
In particular, if $\epsilon=0$ and $\x$ is $k-$sparse, then the recovery is exact.
\end{theorem}

By Theorem \ref{t:2norm} and Lemma \ref{l:hp}, we have the following result.

\begin{corollary} \label{c:2norm}
Suppose that $\A$ and $\e$ in \eqref{e:model} respectively satisfy the RIP with given $\delta_{2k}\in[\frac{\sqrt{2}}{2}, 1)$ and $\|\e\|_2\leq \epsilon$,
then for each $p\in(0,\bar{p}]$, where $\bar{p}$ is defined as in \eqref{e:pbar}, the solution $\x^{\star}$ to the $l_p-$minimization problem \eqref{e:lp} obeys \eqref{e:2norm}. In particular, if $\epsilon=0$ and $\x$ is $k-$sparse, then the recovery is exact.
\end{corollary}

\subsection{Special Case: $n\leq4k$}

In the previous subsection, we have obtained some sufficient conditions to grantee the stably recovery or exactly recovery of the $k-$sparse signal $\x$ from \eqref{e:model} via solving \eqref{e:lp}.
In the following, we will show that these conditions can be further improved under the assumption that $n\leq4k$.
Like in \cite{Mol11}, for simplicity, we assume that $l=3$ (i.e., $n=4k$) throughout this case.

Let
\begin{align}
\label{e:Cpbar}
\bar{C}(p)=(1+\delta_{2k})2^{\frac{p}{2}-1}(\frac{g(p)}{1-\delta_{2k}})^{p/2},
\end{align}
where $g(p)$ is defined as in \eqref{e:gp}. Then we have the following result whose proof will be provided in Section \ref{s:34}.

\begin{theorem} \label{t:pnormbar}
Suppose that $\A$ and $\e$ in \eqref{e:model} respectively satisfy the RIP with given $\delta_{2k}\in[\frac{\sqrt{2}}{2}, 1)$ and $\|\e\|_2\leq \epsilon$, then for each $p\in(0,1)$ such that
\begin{align}
\label{e:Cpbar1}
\bar{C}(p)<1,
\end{align}
the solution $\x^{\star}$ to the $l_p-$minimization problem \eqref{e:lp} obeys
\begin{eqnarray}
\label{e:pnormbar}
\|\x-\x^{\star}\|_p^p\leq \bar{C}_0\|\x_{T_0^c}\|_p^p+\bar{C}_1k^{1-\frac{p}{2}}\epsilon^p,
\end{eqnarray}
where
\begin{align*}
\bar{C}_0=\frac{2(1+\bar{C}(p))}{1-\bar{C}(p)},\; \bar{C}_1=\frac{2^{p+2}}{(1-\delta_{2k})^{\frac{p}{2}}(1-\bar{C}(p))}
\end{align*}
with $\bar{C}(p)$ defined as in \eqref{e:Cpbar}.  In particular, if $\epsilon=0$ and $\x$ is $k-$sparse, then the recovery is exact.
\end{theorem}

Like in the previous subsection, for a given $\delta_{2k}\in[\frac{\sqrt{2}}{2}, 1)$, it may be complicate to determine the range of $p$ such that \eqref{e:Cpbar1} holds. So in the following, we want to give a simple method to determine it. But we first need to give the following lemma whose proof will be provided in Appendix \ref{s:lem2}.

\begin{lemma} \label{l:h1pbar2}
For each given $\delta_{2k}\in[\frac{\sqrt{2}}{2}, 1)$, for each $p\in(0,1)$, if
\beqnn
\delta_{2k}\leq-(6-4\sqrt{2})p+1,
\eeqnn
then \eqref{e:Cpbar1} holds.
\end{lemma}

By Lemma \ref{l:h1pbar2}, for any given $\delta_{2k}\in[\frac{\sqrt{2}}{2}, 1)$,
\eqref{e:Cpbar1} hold for all $p\in(0, \frac{1}{6-4\sqrt{2}}(1-\delta_{2k})]$. Therefore, by Theorem \ref{t:pnormbar}, we immediately have the following corollary.

\begin{corollary} \label{c:pnormbar}
Suppose that $\A$ and $\e$ in \eqref{e:model} respectively satisfy the RIP with given $\delta_{2k}\in[\frac{\sqrt{2}}{2}, 1)$ and $\|\e\|_2\leq \epsilon$, then for each $p\in(0, \frac{3+2\sqrt{2}}{2}(1-\delta_{2k})]$, the solution $\x^{\star}$ to the $l_p-$minimization problem \eqref{e:lp} obeys \eqref{e:pnormbar}. In particular, if $\epsilon=0$ and $\x$ is $k-$sparse, then the recovery is exact.
\end{corollary}

Like in the previous subsection, we also want to bound the $2-$norm of the error.
Let
\begin{eqnarray}
\label{e:Dpbar}
\bar{D}(p)=(\frac{2g(p)}{1-\delta_{2k}})^{p/2},
\end{eqnarray}
where $g(p)$ is defined as in \eqref{e:gp}. Then similarly, we have the following Theorem whose proof will be provided in Section \ref{s:34}.

\begin{theorem} \label{t:2normbar}
Suppose that $\A$ and $\e$ in \eqref{e:model} respectively satisfy the RIP with given $\delta_{2k}\in[\frac{\sqrt{2}}{2}, 1)$ and $\|\e\|_2\leq \epsilon$, then for each $p\in(0,1)$ such that \eqref{e:Cpbar1} holds,
the solution $\x^{\star}$ to the $l_p-$minimization problem \eqref{e:lp} obeys
\begin{eqnarray}
\label{e:2normbar}
\|\x-\x^{\star}\|_2^p\leq \bar{D}_0k^{\frac{p}{2}-1}\|\x_{T_0^c}\|_p^p+\bar{D}_1\epsilon^p,
\end{eqnarray}
where
\begin{align*}
\bar{D}_0=\frac{2\bar{D}(p)}{1-\bar{C}(p)},\,\bar{D}_1=\frac{2^p}{(1-\delta_{2k})^{\frac{p}{2}}}(1+\frac{2\bar{D}(p)}{1-\bar{C}(p)})
\end{align*}
with $\bar{C}(p)$ and $\bar{D}(p)$ defined as in \eqref{e:Cpbar} and \eqref{e:Dpbar}, respectively.
In particular, if $\epsilon=0$ and $\x$ is $k-$sparse, then the recovery is exact.
\end{theorem}

By Theorem \ref{t:2normbar} and Lemma \ref{l:h1pbar2}, we immediately have the following corollary.
\begin{corollary} \label{c:2normbar}
Suppose that $\A$ and $\e$ in \eqref{e:model} respectively satisfy the RIP with given $\delta_{2k}\in[\frac{\sqrt{2}}{2}, 1)$ and $\|\e\|_2\leq \epsilon$, then for each $p\in(0, \frac{3+2\sqrt{2}}{2}(1-\delta_{2k})]$, the solution $\x^{\star}$ to the $l_p-$minimization problem \eqref{e:lp} obeys \eqref{e:2normbar}. In particular, if $\epsilon=0$ and $\x$ is $k-$sparse, then the recovery is exact.
\end{corollary}

\section{Proofs}

In this section, we will prove our main results.
From now on, we always assume that
\beq
\label{e:t}
\|\h_{T_1}\|_p^p=t\|\h_{T_0^c}\|_p^p
\eeq
for some $t\in[0,1]$.


To prove our theorems, we need to use Lemmas \ref{l:01}, \ref{l:pnorm},  \ref{l:h2square} and \ref{l:holder} which were given in \cite{Mol11}, \cite{Sun12}, \cite{LaiL11} and \cite{LaiL11},respectively.

\begin{lemma} \label{l:01}
For $\forall \delta_{2k}\in (0,1)$, it holds that $\|\h_{T_0}\|_2^2+\|\h_{T_1}\|_2^2\leq\frac{1}{1-\delta_{2k}}(2\epsilon+\|\sum_{i=2}^l\A\h_{T_i}\|_2)^2.$
\end{lemma}

%
%

\begin{lemma} \label{l:pnorm}
For $\forall p\in(0,1)$, it holds that $\|\h_{T_0^c}\|_p^p \leq \|\h_{T_0}\|_p^p+2\|\x_{T_0^c}\|_p^p$.
\end{lemma}

\begin{lemma} \label{l:h2square}
For $\forall p\in(0,1)$, it holds that $\sum_{i=2}^l\|\h_{T_i}\|_2^2\leq (1-t)t^{\frac{2}{p}-1}k^{1-\frac{2}{p}}\|\h_{T_0^c}\|_p^2.$
\end{lemma}


\begin{lemma} \label{l:holder}
Let $\u$ be a $k$ dimensional column vector, then for each $p\in(0,1)$, we have $\|\u\|_2\geq k^{\frac{1}{2}-\frac{1}{p}}\|\u\|_p.$
\end{lemma}



\subsection{Proof of Theorem \ref{t:pnorm} and Theorem \ref{t:2norm}} \label{s:12}

Before processing to prove Theorem \ref{t:pnorm} and Theorem \ref{t:2norm}, we need to introduce the following lemmas, where
the proof of Lemma \ref{l:sumsquare} is provided in section Appendix \ref{s:sumsquare}.

\begin{lemma} \label{l:sumsquare}
For $\forall p\in(0,1)$, it holds that $\|\sum_{i=2}^l\A\h_{T_i}\|_2^2\leq C_1(t,p)k^{1-\frac{2}{p}} \|\h_{T_0^c}\|_p^2,$
where
\begin{eqnarray}
\label{e:C1tp}
C_1(t,p)=
\begin{cases}
(1-t)t^{\frac{2}{p}-1}+2g(p)\delta_{2k}, &p\in(0,p^{\star}]\cr
(1-t)t^{\frac{2}{p}-1}+2^{2-\frac{2}{p}}\delta_{2k}, &p\in(p^{\star},1)
\end{cases}
\end{eqnarray}
 with $g(p)$ defined as in \eqref{e:gp}.
\end{lemma}

\begin{remark}
The upper bound on $\|\sum_{i=2}^l\A\h_{T_i}\|_2^2$ given by Lemma \ref{l:sumsquare} is sharper than that of the Lemma 5 in \cite{Wuc13},
where the bound is $((1-t)t^{\frac{2}{p}-1}+2\delta_{2k}p(1-\frac{p}{2})^{\frac{2}{p}-1})k^{1-\frac{2}{p}} \|\h_{T_0^c}\|_p^2$. In fact, to show
this, it suffices to show the following inequality:
\begin{eqnarray*}
\begin{cases}
g(p)\leq p(1-\frac{p}{2})^{\frac{2}{p}-1}, &p\in(0,p^{\star}]\cr
2^{1-\frac{2}{p}}\leq p(1-\frac{p}{2})^{\frac{2}{p}-1}, &p\in(p^{\star},1)
\end{cases}.
\end{eqnarray*}
It is not hard to check that the aforementioned inequality follows from \eqref{e:gp} and \eqref{e:fpp}.
\end{remark}

\begin{lemma} \label{l:0}
For $\forall \delta_{2k}\in [\frac{\sqrt{2}}{2},1)$, for each $p\in(0,1)$, it holds that
\beqnn
\|\h_{T_0}\|_p^p \leq  \frac{2^{\frac{3p}{2}}}{(1-\delta_{2k})^{\frac{p}{2}}}
k^{1-\frac{p}{2}}\epsilon^p+C(p)\|\h_{T_0^c}\|_p^p,
\eeqnn
where $C(p)$ is defined as in \eqref{e:Cp}.
\end{lemma}

{\em Proof}. By Lemma \ref{l:01} and Lemma \ref{l:sumsquare}
\begin{eqnarray}
\label{e:h0square}
&(1-\delta_{2k})\|\h_{T_0}\|_2^2 \leq4\epsilon^2+4\epsilon\|\sum_{i=2}^l\A\h_{T_i}\|_2+\|\sum_{i=2}^l\A\h_{T_i}\|_2^2-(1-\delta_{2k})\|\h_{T_1}\|_2^2,\\
\label{e:k1}
&k^{\frac{2}{p}-1}\|\sum_{i=2}^l\A\h_{T_i}\|_2^2\leq C_1(t,p)\|\h_{T_0^c}\|_p^2.
\end{eqnarray}
where $C_1(t,p)$ is defined as in \eqref{e:C1tp}.
By \eqref{e:t} and Lemma \ref{l:holder}
\begin{align*}
k^{\frac{2}{p}-1}\|\h_{T_1}\|_2^2\geq  t^{\frac{2}{p}}\|\h_{T_0^c}\|_p^2.
\end{align*}
By\eqref{e:C1tp}, \eqref{e:k1} and the aforementioned inequality, we have
\begin{align}
\label{e:k2}
k^{\frac{2}{p}-1}(\|\sum_{i=2}^l\A\h_{T_i}\|_2^2-(1-\delta_{2k})\|\h_{T_1}\|_2^2) \leq C_2(t,p)\|\h_{T_0^c}\|_p^2,
\end{align}
where
\begin{eqnarray*}
\label{e:C2tp}
&C_2(t,p)=
\begin{cases}
(\delta_{2k}-2)t^{\frac{2}{p}}+t^{\frac{2}{p}-1}+2g(p)\delta_{2k}, &p\in(0,p^{\star}]\cr
(\delta_{2k}-2)t^{\frac{2}{p}}+t^{\frac{2}{p}-1}+2^{2-\frac{2}{p}}\delta_{2k}, &p\in(p^{\star},1)
\end{cases}.
\end{eqnarray*}

By \eqref{e:gp} and \eqref{e:C1tp}, one can easily show that
\begin{eqnarray}
\label{e:C1max}
&C_1(t,p)\leq C_1(1-\frac{p}{2},p)=
\begin{cases}
(1+2\delta_{2k})g(p), &p\in(0,p^{\star}]\cr
g(p)+2^{2-\frac{2}{p}}\delta_{2k}, &p\in(p^{\star},1)
\end{cases};\\
\label{e:C2max}
&C_2(t,p)\leq C_2(\frac{2-p}{2(2-\delta_{2k})},p)=
\begin{cases}
((2-\delta_{2k})^{1-\frac{2}{p}}+2\delta_{2k})g(p), &p\in(0,p^{\star}]\cr
(2-\delta_{2k})^{1-\frac{2}{p}}g(p)+2^{2-\frac{2}{p}}\delta_{2k}, &p\in(p^{\star},1)
\end{cases}.
\end{eqnarray}

By \eqref{e:gpr} and \eqref{e:pstar} , for each $\delta_{2k}\in [\frac{\sqrt{2}}{2},1)$ and $p\in(0,1)$, we have
$g(p)\leq 2\delta_{2k}g(p)$ and for each $p\in(p^{\star},1)$, we have $g(p)\leq 2^{2-\frac{2}{p}}\delta_{2k}$.
Therefore, by \eqref{e:C1max} and \eqref{e:C2max}, we have
\beqnn
C_1(1-\frac{p}{2},p)\leq 2C_2(\frac{2-p}{2(2-\delta_{2k})},p).
\eeqnn

By the aforementioned equation, \eqref{e:k1} and \eqref{e:k2}, we have
\begin{align*}
k^{\frac{1}{p}-\frac{1}{2}}\|\sum_{i=2}^l\A\h_{T_i}\|_2\leq \sqrt{2}\sqrt{C_2(\frac{2-p}{2(2-\delta_{2k})},p)}\|\h_{T_0^c}\|_p;
\end{align*}
\begin{align*}
k^{\frac{2}{p}-1}(\|\sum_{i=2}^l\A\h_{T_i}\|_2^2-(1-\delta_{2k})\|\h_{T_1}\|_2^2)
\leq C_2(\frac{2-p}{2(2-\delta_{2k})},p)\|\h_{T_0^c}\|_p^2.
\end{align*}

By the aforementioned two inequalities and \eqref{e:h0square},
\begin{align*}
(1-\delta_{2k})k^{\frac{2}{p}-1}\|\h_{T_0}\|_2^2 \leq (2^{\frac{3}{2}}k^{\frac{1}{p}-\frac{1}{2}}\epsilon+\sqrt{C_2(\frac{2-p}{2(2-\delta_{2k})},p)}\|\h_{T_0^c}\|_p)^2.
\end{align*}

By \eqref{e:Cp}, \eqref{e:C2max} and applying Lemma \ref{l:holder} to the aforementioned inequality, we have,
\beqnn
\|\h_{T_0}\|_p \leq \frac{2^{\frac{3}{2}}}{\sqrt{1-\delta_{2k}}}k^{\frac{1}{p}-\frac{1}{2}}\epsilon+(C(p))^{\frac{1}{p}}\|\h_{T_0^c}\|_p.
\eeqnn
The lemma follows from the aforementioned inequality and the fact that for each fixed $n\in N$, for each $\omega_j\geq0, 1\leq j\leq n$ and for each $p\in (0,1]$, it holds that
\beq
\label{e:omega}
\sum_{j=1}^n\omega_j\leq(\sum_{j=1}^n\omega_j^p)^{\frac{1}{p}}.
\eeq
In fact, if $(\sum_{j=1}^n\omega_j^p)^{\frac{1}{p}}=0$, then \eqref{e:omega} obviously holds. Otherwise,
we assume $u_j=\frac{\omega_j}{(\sum_{j=1}^n\omega_j^p)^{\frac{1}{p}}}$. Then $u_j\leq1$ and
$\sum_{j=1}^nu_j^p=1$. Since $p\in (0,1)$, we have $\sum_{j=1}^nu_j\leq\sum_{j=1}^nu_j^p=1$,
\ \ $\Box$

{\em Proof of Theorem \ref{t:pnorm}}. By Lemma \ref{l:pnorm}, we have,
\begin{align*}
\|\h\|_p^p=\|\h_{T_0}\|_p^p+\|\h_{T_0^c}\|_p^p\leq 2\|\h_{T_0}\|_p^p+2\|\x_{T_0^c}\|_p^p.
\end{align*}
If \eqref{e:Cp1} holds, then by lemmas \ref{l:pnorm} and \ref{l:0}, we have
\begin{align*}
\|\h_{T_0}\|_p^p&\leq \frac{2C(p)}{1-C(p)}\|\x_{T_0^c}\|_p^p+
\frac{2^{\frac{3p}{2}}}{(1-\delta_{2k})^{\frac{p}{2}}}\frac{k^{1-\frac{p}{2}}}{1-C(p)}\epsilon^p.
\end{align*}
The aforementioned two equations imply the theorem.


{\em Proof of Theorem \ref{t:2norm}}. By Lemma \ref{l:01}, we have
\begin{align*}
&\|\h\|_2^2=(\|\h_{T_0}\|_2^2+\|\h_{T_1}\|_2^2)+\sum_{i=2}^l\|\h_{T_i}\|_2^2
\leq\frac{1}{1-\delta_{2k}}(2\epsilon+\|\sum_{i=2}^l\A\h_{T_i}\|_2)^2+\sum_{i=2}^l\|\h_{T_i}\|_2^2.
\end{align*}
Hence,
\begin{align}
\label{e:h2square2}
\|\h\|_2\leq\frac{2\epsilon}{\sqrt{1-\delta_{2k}}}+ \sqrt{\frac{1}{1-\delta_{2k}}\|\sum_{i=2}^l\A\h_{T_i}\|_2^2+\sum_{i=2}^l\|\h_{T_i}\|_2^2}.
\end{align}
Then by lemmas \ref{l:h2square} and \ref{l:sumsquare}, we have
\begin{align*}
\|\h\|_2\leq\frac{2\epsilon}{\sqrt{1-\delta_{2k}}}+ \frac{k^{\frac{1}{2}-\frac{1}{p}}}{\sqrt{1-\delta_{2k}}}\sqrt{C_3(t,p)}\|\h_{T_0^c}\|_p,
\end{align*}
where
\begin{eqnarray*}
C_3(t,p)=
\begin{cases}
(2-\delta_{2k})(1-t)t^{\frac{2}{p}-1}+2g(p)\delta_{2k}, &p\in(0,p^{\star}]\cr
(2-\delta_{2k})(1-t)t^{\frac{2}{p}-1}+2^{2-\frac{2}{p}}\delta_{2k}, &p\in(p^{\star},1)
\end{cases}.
\end{eqnarray*}
Therefore, by \eqref{e:omega}, we have,
\begin{align*}
\|\h\|_2^p\leq\frac{2^p}{(1-\delta_{2k})^{\frac{p}{2}}}\epsilon^p+ (\frac{C_3(t,p)}{1-\delta_{2k}})^{\frac{p}{2}}k^{\frac{p}{2}-1}\|\h_{T_0^c}\|_p^p.
\end{align*}
By some simple calculations, for each $t\in [0,1]$, we have
\beqnn
(\frac{C_3(t,p)}{1-\delta_{2k}})^{\frac{p}{2}}\leq D(p),
\eeqnn
where $D(p)$ is defined as in \eqref{e:Dp}, so
\begin{align*}
\|\h\|_2^p\leq\frac{2^p}{(1-\delta_{2k})^{\frac{p}{2}}}\epsilon^p+ D(p)k^{\frac{p}{2}-1}\|\h_{T_0^c}\|_p^p.
\end{align*}
If \eqref{e:Cp1} holds, then by lemmas \ref{l:pnorm} and \ref{l:0}, we have
\begin{align*}
\|\h_{T_0^c}\|_p^p&\leq \frac{2}{1-C(p)}\|\x_{T_0^c}\|_p^p+ \frac{2^{\frac{3p}{2}}}{(1-\delta_{2k})^{\frac{p}{2}}}\frac{k^{1-\frac{p}{2}}}{1-C(p)}\epsilon^p.
\end{align*}
The aforementioned two equations imply the theorem.

\subsection{Proof of Theorem \ref{t:pnormbar} and Theorem \ref{t:2normbar} } \label{s:34}

In this subsection, we will follow the method used in \cite{Mol11} to prove Theorem \ref{t:pnormbar} and Theorem \ref{t:2normbar}. But before proving them, we need to introduce the following lemma.

\begin{lemma} \label{l:02}
For each $\delta_{2k}\in [\frac{\sqrt{2}}{2},1)$, for each $p\in(0,1)$,
\beqnn
\|\h_{T_0}\|_p^p \leq
\frac{2^{p+1}}{(1-\delta_{2k})^{\frac{p}{2}}}k^{1-\frac{p}{2}}\epsilon^p+\bar{C}(p)\|\h_{T_0^c}\|_p^p,
\eeqnn
where $\bar{C}(p)$ is defined as in \eqref{e:Cpbar}.
\end{lemma}
{\em Proof}.
By Lemma \ref{l:h2square},
\begin{align}
\label{e:k12}
&k^{\frac{1}{p}-\frac{1}{2}}\|\sum_{i=2}^3\A\h_{T_i}\|_2\leq k^{\frac{1}{p}-\frac{1}{2}}\sqrt{(1+\delta_{2k})\sum_{i=2}^3\|\h_{T_i}\|_2^2} \nonumber\\
&\leq \sqrt{1+\delta_{2k}}\sqrt{(1-t)t^{\frac{2}{p}-1}}\|\h_{T_0^c}\|_p
\leq\sqrt{1+\delta_{2k}}\sqrt{g(p)}\|\h_{T_0^c}\|_p,
\end{align}
where $g(p)$ is defined as in \eqref{e:gp}.

By the aforementioned inequality, \eqref{e:t} and Lemma \ref{l:holder}, we have
\begin{align}
\label{e:k22}
&k^{\frac{2}{p}-1}(\|\sum_{i=2}^3\A\h_{T_i}\|_2^2-(1-\delta_{2k})\|\h_{T_1}\|_2^2)
\leq [(1+\delta_{2k})(1-t)t^{\frac{2}{p}-1}-(1-\delta_{2k})t^{\frac{2}{p}}])\|\h_{T_0^c}\|_p^2\nonumber\\
&=(1+\delta_{2k}-2t)t^{\frac{2}{p}-1}\|\h_{T_0^c}\|_p^2
\leq (1+\delta_{2k})^{\frac{2}{p}}2^{1-\frac{2}{p}}g(p)\|\h_{T_0^c}\|_p^2.
\end{align}
Obviously, for each $\delta_{2k}\in [\frac{\sqrt{2}}{2},1)$, for each $p\in(0,1)$,
\beqnn
2^{\frac{1}{p}}\times2^{\frac{1}{2}-\frac{1}{p}}(1+\delta_{2k})^{\frac{1}{p}}\geq\sqrt{1+\delta_{2k}}.
\eeqnn
By the aforementioned inequalities, \eqref{e:h0square} and \eqref{e:k12}, we have
\begin{align*}
(1-\delta_{2k})k^{\frac{2}{p}-1}\|\h_{T_0}\|_2^2 \leq (2^{1+\frac{1}{p}}k^{\frac{1}{p}-\frac{1}{2}}\epsilon+(1+\delta_{2k})^{\frac{1}{p}}2^{\frac{1}{2}-\frac{1}{p}}\sqrt{g(p)}\|\h_{T_0^c}\|_p)^2.
\end{align*}
By \eqref{e:Cpbar} and applying Lemma \ref{l:holder} to the aforementioned inequality, we have,
\beqnn
\|\h_{T_0}\|_p \leq \frac{2^{1+\frac{1}{p}}}{\sqrt{1-\delta_{2k}}}k^{\frac{1}{p}-\frac{1}{2}}\epsilon+(\bar{C}(p))^{\frac{1}{p}}\|\h_{T_0^c}\|_p.
\eeqnn
By \eqref{e:omega} and the aforementioned inequality, the lemma holds.
\ \ $\Box$


{\em Proof of Theorem \ref{t:pnormbar}}. By Lemma \ref{l:pnorm}, we have
\begin{align*}
\|\h\|_p^p=\|\h_{T_0}\|_p^p+\|\h_{T_0^c}\|_p^p\leq 2\|\h_{T_0}\|_p^p+2\|\x_{T_0^c}\|_p^p.
\end{align*}

If \eqref{e:Cpbar1} holds, then by Lemma \ref{l:pnorm} and \ref{l:02}, we have
\begin{align*}
\|\h_{T_0}\|_p^p&\leq \frac{2\bar{C}(p)}{(1-\bar{C}(p))}\|\x_{T_0^c}\|_p^p+\frac{2^{p+1}k^{1-\frac{p}{2}}}{(1-\delta_{2k})^{\frac{p}{2}}(1-\bar{C}(p))}\epsilon^p.
\end{align*}

The aforementioned two equations imply the theorem.

{\em Proof of Theorem \ref{t:2normbar}}. By \eqref{e:h2square2}, \eqref{e:k12} and Lemma \ref{l:h2square}, we have
\begin{align*}
\|\h\|_2&\leq\frac{2\epsilon}{\sqrt{1-\delta_{2k}}}+ \frac{k^{\frac{1}{2}-\frac{1}{p}}}{\sqrt{1-\delta_{2k}}}\sqrt{2(1-t)t^{\frac{2}{p}-1}}\|\h_{T_0^c}\|_p\leq\frac{2\epsilon}{\sqrt{1-\delta_{2k}}}+ \frac{k^{\frac{1}{2}-\frac{1}{p}}}{\sqrt{1-\delta_{2k}}}\sqrt{2g(p)}\|\h_{T_0^c}\|_p.
\end{align*}
Therefore, by \eqref{e:Dpbar} and \eqref{e:omega}, we have
\begin{align*}
\|\h\|_2^p\leq\frac{2^p}{(1-\delta_{2k})^{\frac{p}{2}}}
\epsilon^p+ \bar{D}(p)k^{\frac{p}{2}-1}\|\h_{T_0^c}\|_p^p.
\end{align*}
If \eqref{e:Cpbar1} holds, then by Lemma \ref{l:pnorm} and \ref{l:02}, we have
\begin{align*}
\|\h_{T_0^c}\|_p^p&\leq \frac{2}{1-\bar{C}(p)}\|\x_{T_0^c}\|_p^p+ \frac{2^{p+1}k^{1-\frac{p}{2}}}{(1-\delta_{2k})^{\frac{p}{2}}(1-\bar{C}(p))}\epsilon^p.
\end{align*}
where $\bar{C}(p)$ is defined as in \eqref{e:Cpbar}. The aforementioned two equations imply the theorem.

\section{Summary and future work} \label{s:con}
In this paper, we showed that, under the assumption that $\|\e\|_2\leq \epsilon$, every $k-$sparse signal $\x\in \mathbb{R}^n$ can be stably ($\epsilon\neq0$) or exactly recovered ($\epsilon=0$) from \eqref{e:model} via $l_p-$mnimization with $p\in(0, \bar{p}]$, where $\bar{p}$ is defined as in \eqref{e:pbar}, even if $\delta_{2k}\in[\frac{\sqrt{2}}{2}, 1)$.
Furthermore, under the assumption that $n\leq 4k$, we showed that the range of $p$ can be further improved to $p\in(0,\frac{3+2\sqrt{2}}{2}(1-\delta_{2k})]$. This not only extended some discussions of only the noiseless recovery \cite{LaiL11}, \cite{Wuc13} to the noise recovery, but also greatly improves the best existing results where $p<\min\{1, 1.0873(1-\delta_{2k}) \}$ \cite{Wuc13}.

In the future, we will discuss the largest possible ranges of $p$ for a given $\delta_{2k}\in[\frac{\sqrt{2}}{2}, 1)$ and how to chose $p$ if $\delta_{2k}$ is not given.


\section*{Acknowledgment}
This work is supported in part by NSFC (Grant No. 11201161, 11171125, 91130003, 71171168).

\appendices
\section{Proof of Lemma \ref{l:hp} } \label{s:lem1}

Before proving Lemma \ref{l:hp}, we need to give the following lemma.
\begin{lemma} \label{l:phi}
For each $t\in(0,1)$, let
\beq
\label{e:phi12}
\phi_1(t)=(\frac{1-t}{1+t})^{\frac{1}{t}-1}\mbox{ and } \phi_2(t)=(1-t)^{\frac{1}{t}}.
\eeq
Then $\phi_1$ and $\phi_2$ are respectively monotonically increasing and decreasing functions.
\end{lemma}

{\em Proof. } 
Since
\beqnn
(\ln\phi_1(t))'=-\frac{2}{t(1+t)}-\frac{1}{t^2}\ln(\frac{1-t}{1+t})\mbox{ and }  (\ln\phi_2(t))'=-\frac{1}{t(1-t)}-\frac{1}{t^2}\ln(1-t),
\eeqnn
it is equivalent to show
\beqnn
\ln(\frac{1+t}{1-t})>\frac{2t}{1+t} \mbox{ and } -\ln(1-t)<\frac{t}{1-t}.
\eeqnn
One can easily show that the above inequalities hold for each $t\in(0,1)$. \ \ $\Box$

%

In the following, we will prove Lemma \ref{l:hp}.

{\em Proof of Lemma \ref{l:hp}}. Firstly, we prove the lemma holds for $p\in(0,p^{\star}]$.
Obviously, it suffices to show
\beqnn
((2-\delta_{2k})^{1-\frac{2}{p}}+2\delta_{2k})g(p)+\delta_{2k}<1.
\eeqnn
By direct calculation and \eqref{e:gpr}, for each fixed $p$, the left hand side of the above inequality is a monotonically increasing function of $\delta_{2k}$, so it suffices to show
\beqnn
[(1+\frac{p}{2})^{1-\frac{2}{p}}+2(1-\frac{p}{2})]g(p)<\frac{p}{2}.
\eeqnn
By \eqref{e:gp} and \eqref{e:phi12}, it is equivalent to show
\beqnn
\phi_1(\frac{p}{2})+2\phi_2(\frac{p}{2})<1.
\eeqnn
By \eqref{e:pstar} and Lemma \ref{l:phi}, for each $p\in(0,p^{\star}]$, we have
\beqnn
\phi_1(\frac{p}{2})+2\phi_2(\frac{p}{2})\leq\phi_1(0.25)+2\lim_{p\rightarrow0^+}\phi_2(p)<1,
\eeqnn
so the lemma holds in this case.

Secondly, we prove the lemma holds for $p\in(p^{\star},1)$.
Similarly, it suffices to show
\beqnn
(1+0.62p)^{1-\frac{2}{p}}g(p)+2^{2-\frac{2}{p}}(1-0.62p)<0.62p.
\eeqnn
By \eqref{e:gp}, we only need to show
\beq
\label{e:2p}
(1+0.62p)^{1-\frac{2}{p}}(1-\frac{p}{2})^{\frac{2}{p}-1}+2^{3-\frac{2}{p}}(\frac{1}{p}-0.62)<1.24.
\eeq
By \eqref{e:phi12} and Lemma \ref{l:phi}, it is easy to check that for each $p\in(p^{\star},1)$, we have
\beqnn
(1+0.62p)^{1-\frac{2}{p}}(1-\frac{p}{2})^{\frac{2}{p}-1}<\phi_1(\frac{1}{2})=\frac{1}{3}.
\eeqnn
It is easy to verify that $2^{3-\frac{2}{p}}(\frac{1}{p}-0.62)$ achieves the maximal value at
$\frac{1}{p}=0.62+\frac{1}{2\ln(2)}$, therefore for each $p\in(p^{\star},1)$, we have
\beqnn
2^{3-\frac{2}{p}}(\frac{1}{p}-0.62)<2^{1.76-\frac{1}{\ln(2)}}\frac{1}{2\ln(2)}<0.8988.
\eeqnn
By the aforementioned two inequations, \eqref{e:2p} holds and this finishes the proof.
\ \ $\Box$

\section{Proof of Lemma \ref{l:h1pbar2} } \label{s:lem2}

Before proving Lemma \ref{l:h1pbar2}, we need to give the following lemma.
\begin{lemma} \label{l:varphi}
For each $p\in(0,1]$, let
\beq
\label{e:varphi}
\varphi(p)=(1-(3-2\sqrt{2})p)^{\frac{2}{p}}(1-\frac{p}{2})^{\frac{2}{p}-1},
\eeq
then $\varphi(p)$ is a monotonically increasing function on  $(0,1]$.
\end{lemma}

{\em Proof } By some simple calculations, we have
\begin{align*}
(\ln(\varphi(p)))'=-\frac{1}{p}[\frac{2(3-2\sqrt{2})}{1-(3-2\sqrt{2})p}+1]
-\frac{2}{p^2}[\ln(1-(3-2\sqrt{2})p)+\ln(1-\frac{p}{2})].
\end{align*}
So it suffices to show, for each $p\in(0,1]$,
\beqnn
\bar{\varphi}(p)=\frac{(3-2\sqrt{2})p}{1-(3-2\sqrt{2})p}+\frac{p}{2}+\ln(1-(3-2\sqrt{2})p)+\ln(1-\frac{p}{2})\leq0.
\eeqnn
One can easily show that for each $p\in(0,1]$,
\begin{align*}
\bar{\varphi}'(p)=\frac{(3-2\sqrt{2})}{[1-(3-2\sqrt{2})p]^2}-\frac{(3-2\sqrt{2})}{1-(3-2\sqrt{2})p}+\frac{1}{2}
-\frac{1}{2-p}=\frac{(3-2\sqrt{2})^2p}{[1-(3-2\sqrt{2})p]^2}+\frac{1}{2}-\frac{1}{2-p}\leq0.
\end{align*}
Therefore $\bar{\varphi}(p)\leq\bar{\varphi}(0)=0.$ Thus the lemma is proved.\ \ $\Box$

{\em Proof of Lemma \ref{l:h1pbar2}}. By  \eqref{e:Cpbar} and \eqref{e:Cpbar1}, obviously, it suffices to show
\beqnn
(1+\delta_{2k})^{\frac{2}{p}}2^{1-\frac{2}{p}}g(p)+\delta_{2k}<1.
\eeqnn
By \eqref{e:gpr}, the left hand side of the above inequality is a monotonically increasing function of $\delta_{2k}$, so it suffices to show
\beqnn
(2-(6-4\sqrt{2})p)^{\frac{2}{p}}2^{1-\frac{2}{p}}g(p)<(6-4\sqrt{2})p.
\eeqnn
By \eqref{e:gp}, we only need to show
\beqnn
(1-(3-2\sqrt{2})p)^{\frac{2}{p}}(1-\frac{p}{2})^{\frac{2}{p}-1}<(6-4\sqrt{2}).
\eeqnn
By Lemma \ref{l:varphi}, for each $p\in(0,1)$, we have
\beqnn
(1-(3-2\sqrt{2})p)^{\frac{2}{p}}(1-\frac{p}{2})^{\frac{2}{p}-1}<\varphi(1) =(6-4\sqrt{2}),
\eeqnn
so the lemma holds.
\ \ $\Box$

\section{Proof of Lemma \ref{l:sumsquare} } \label{s:sumsquare}

Before proving Lemma \ref{l:sumsquare} we need to introduce the following lemma whose proof will be provided in the latter part of this subsection.
\begin{lemma} \label{l:reverse}
For $\forall p\in(0,1)$, it holds that,
\beqnn
\sum_{i=2}^l\|\h_{T_i}\|_2\leq \sqrt{2}C_1(p)k^{\frac{1}{2}-\frac{1}{p}}\|\h_{T_0^c}\|_p,
\eeqnn
where \begin{eqnarray}
\label{e:C1p}
C_1(p)=
\begin{cases}
(\frac{p}{2})^{\frac{1}{2}}(\frac{2}{2-p})^{\frac{1}{2}-\frac{1}{p}}, &p\in(0,p^{\star}]\cr
2^{\frac{1}{2}-\frac{1}{p}}, &p\in(p^{\star},1)
\end{cases}.
\end{eqnarray}
\end{lemma}

\begin{remark}
The bound given by Lemma \ref{l:reverse} is sharper than the corresponding bound
given in \cite{LaiL11} (lemma 4), in \cite{Wuc13} (lemma 4) and in \cite{BahR13} (lemma 2.4). To save the space, we do not give the details.
%
\end{remark}

{\em Proof of Lemma \ref{l:sumsquare}}.
For each $i, j\geq 1$ and $i\neq j$, $T_i\bigcap T_j=\phi$, therefore, by Lemma 2.1 in \cite{Can08}, we have
\beqnn
|\langle\A\h_{T_i}, \A\h_{T_j}\rangle|\leq\delta_{2k}\|\h_{T_i}\|_2\|\h_{T_j}\|_2.
\eeqnn
By the aforementioned equation and \eqref{e:RIP}, we have,
\begin{align*}
&\|\sum_{i=2}^l\A\h_{T_i}\|_2^2=\sum_{i,j\geq2}^l\langle\A\h_{T_i}, \A\h_{T_j}\rangle
\leq \sum_{i=2}^l|\langle\A\h_{T_i}, \A\h_{T_i}\rangle|+2 \sum_{i>j\geq2}^l|\langle\A\h_{T_i}, \A\h_{T_i}\rangle|\\
&\leq \sum_{i=2}^l(1+\delta_{2k})\|\h_{T_i}\|_2^2+2\delta_{2k} \sum_{i>j\geq2}^l\|\h_{T_i}\|_2\|\h_{T_j}\|_2
= \sum_{i=2}^l\|\h_{T_i}\|_2^2+\delta_{2k}(\sum_{i=2}^l\|\h_{T_i}\|_2)^2.
\end{align*}
So the lemma follows from Lemmas \ref{l:reverse} and \ref{l:h2square}.
\ \ $\Box$


Before following the methods used in \cite{BahR13} and \cite{Wuc13} to prove Lemma \ref{l:reverse}, we introduce the following lemma which was provided in \cite{Fou10}.

\begin{lemma} \label{l:shift}
If $p\in(0,2)$ and $u_1\geq\ldots \geq u_l\geq u_{l+1}\geq\ldots \geq u_r\geq u_{r+1}\geq\ldots \geq u_{r+l}\geq0$, then
\beqnn
(\sum_{i=l+1}^{l+r}u_i^2)^{1/2}\leq C (\sum_{i=1}^{r}u_i^p)^{1/p},
\eeqnn
where $C=\max\{r^{\frac{1}{2}-\frac{1}{p}},(\frac{p}{2})^{\frac{1}{2}}(\frac{2l}{2-p})^{\frac{1}{2}-\frac{1}{p}}\}$.
\end{lemma}

By \eqref{e:fp}, \eqref{e:fpr} and Lemma \ref{l:shift}, we immediately obtain the following corollary.
\begin{corollary} \label{c:shift}
If $p\in(0,1)$ and $u_1\geq\ldots \geq u_k\geq u_{k+1}\geq\ldots \geq u_{2k}\geq u_{2k+1}\geq\ldots \geq u_{3k}\geq0$, then
\beqnn
(\sum^{3k}_{i=k+1}u_i^2)^{1/2}\leq C_1(p)k^{\frac{1}{2}-\frac{1}{p}} (\sum_{i=1}^{2k}u_i^p)^{1/p},
\eeqnn
where $C_1(p)$ is defined as in \eqref{e:C1p}.
\end{corollary}

\begin{remark}
In corollary 1 in \cite{Wuc13}, $C_1(p)=p^{\frac{1}{2}}(\frac{2}{2-p})^{\frac{1}{2}-\frac{1}{p}}$. By \eqref{e:fpp}, $p^{\frac{1}{2}}(\frac{2}{2-p})^{\frac{1}{2}-\frac{1}{p}}\geq 2^{\frac{1}{2}-\frac{1}{p}}$ for each $p\in(p^{\star},1)$, so our bound on $(\sum_{i=k+1}^{3k}u_i^2)^{1/2}$ is sharper.
\end{remark}

{\em Proof of Lemma \ref{l:reverse}}. For every even $j\in\{2,4,\ldots,\}$, obviously, $T_j\bigcap T_{j+1}=\emptyset$. Therefore, one
can easily show that
\beqnn
\|h_{T_j}\|_2+\|h_{T_{j+1}}\|_2\leq \sqrt{2}\|h_{T_j\cup T_{j+1}}\|_2.
\eeqnn
Summing up all the aforementioned inequalities for $j\in{2,4,\ldots,}$ yields
\beqnn
\sum_{j=2}^l\|h_{T_j}\|_2\leq \sqrt{2}\sum_{j=1}\|h_{T_{2j}\cup T_{2j+1}}\|_2.
\eeqnn
Since $|T_j|=k$ for each $j\geq1$. By Corollary \ref{c:shift}, we have,
\beqnn
\|h_{T_{2j}\cup T_{2j+1}}\|_2\leq C_1(p)k^{\frac{1}{2}-\frac{1}{p}}\|h_{T_{2j-1}\cup T_{2j}}\|_p.
\eeqnn
By the aforementioned inequalities, we have
\beqnn
\sum_{j=2}^l\|h_{T_j}\|_2\leq \sqrt{2}C_1(p)k^{\frac{1}{2}-\frac{1}{p}}\sum_{j=1}\|h_{T_{2j-1}\cup T_{2j}}\|_p.
\eeqnn
The lemma follows from the aforementioned equation and \eqref{e:omega}.
\ \ $\Box$

%

\bibliographystyle{IEEEtran}
\bibliography{ref-RIP}

\begin{thebibliography}{10}
\providecommand{\url}[1]{#1}
\csname url@samestyle\endcsname
\providecommand{\newblock}{\relax}
\providecommand{\bibinfo}[2]{#2}
\providecommand{\BIBentrySTDinterwordspacing}{\spaceskip=0pt\relax}
\providecommand{\BIBentryALTinterwordstretchfactor}{4}
\providecommand{\BIBentryALTinterwordspacing}{\spaceskip=\fontdimen2\font plus
\BIBentryALTinterwordstretchfactor\fontdimen3\font minus
  \fontdimen4\font\relax}
\providecommand{\BIBforeignlanguage}[2]{{%
\expandafter\ifx\csname l@#1\endcsname\relax
\typeout{** WARNING: IEEEtran.bst: No hyphenation pattern has been}%
\typeout{** loaded for the language `#1'. Using the pattern for}%
\typeout{** the default language instead.}%
\else
\language=\csname l@#1\endcsname
\fi
#2}}
\providecommand{\BIBdecl}{\relax}
\BIBdecl

\bibitem{CanT05}
E.~J. Cand{\'e}s and T.~Tao, ``Decoding by linear programming,'' \emph{IEEE
  Trans. Inf. Theory}, vol.~51, no.~12, pp. 4203--4215, 2005.

\bibitem{Don06}
D.~L. Donoho, ``Compressed sensing,'' \emph{IEEE Trans. Inf. Theory}, vol.~52,
  no.~4, pp. 1289--1306, 2006.

\bibitem{CohDD09}
A.~Cohen, W.~Dahmen, and R.~DeVore, ``Compressed sensing and best $k$-term
  approximation,'' \emph{J. Amer. Math. Soc.}, vol.~22, pp. 211--231, 2009.

\bibitem{Fuc05}
J.~J. Fuchs., ``Recovery of exact sparse representations in the presence of
  bounded noise,'' \emph{IEEE Trans. Inf. Theory}, vol.~51, no.~10, pp.
  3601--3608, 2005.

\bibitem{DonET06}
D.~L. Donoho, M.~Elad, and V.~N. Temlyakov, ``Stable recovery of sparse
  overcomplete representations in the presence of noise,'' \emph{IEEE Trans.
  Inf. Theory}, vol.~52, pp. 6--18, 2006.

\bibitem{Can08}
E.~J. Cand{\'e}s, ``The restricted isometry property and its implications for
  compressed sensing,'' \emph{C. R. Acad. Sci. Paris, Ser. I}, vol. 346,
  no.~11, pp. 589--592, 2008.

\bibitem{CanRT06}
E.~J. Cand{\'e}s, J.~Romberg, and T.~Tao, ``Stable signal recovery from
  incomplete and inaccurate measurements,'' \emph{Comm. Pure Appl. Math},
  vol.~59, pp. 1207--1223, 2006.

\bibitem{Mol11}
Q.~Mo and S.~Li, ``New bounds on the restricted isometry constant
  $\delta_{2k}$,'' \emph{Appl. Comput. Harmon. Anal.}, vol.~31, pp. 460--468,
  2011.

\bibitem{CanT07}
E.~J. Cand{\'e}s and T.~Tao, ``The dantzig selector: Statistical estimation
  when $p$ is much larger than $n$,'' \emph{Ann. Statist}, vol.~35, pp.
  2313--2351, 2007.

\bibitem{CaiZ13}
T.~Cai and A.~Zhang, ``Sharp {RIP} bound for sparse signal and low-rank matrix
  recovery,'' \emph{Appl. Comput. Harmon. Anal.}, vol.~35, pp. 74--93, 2013.

\bibitem{FouL09}
S.~Foucart and M.-J. Lai, ``Sparsest solutions of underdetermined linear
  systems via $l_q-$ minimization for $0<q\leq1$,'' \emph{Appl. Comput. Harmon.
  Anal.}, vol.~26, no.~3, pp. 395--407, 2009.

\bibitem{Fou10b}
S.~Foucart, ``A note on guaranteed sparse recovery via $l_1-$ minimization,''
  \emph{Appl. Comput. Harmon. Anal.}, vol.~29, pp. 97--103, 2010.

\bibitem{CaiZ13b}
T.~Cai and A.~Zhang, ``Sparse representation of a polytope and recovery of
  sparse signals and low-rank matrices,'' \emph{IEEE Trans. Inf. Theory},
  vol.~60, no.~1, pp. 122--132.

\bibitem{DavG09}
M.~E. Davies and R.~Gribonval, ``Restricted isometry constants where $l^p$
  sparse recovery can fail for $0<p\leq1$,'' \emph{IEEE Trans. Inf. Theory},
  vol.~55, no.~5, pp. 2203--2214, 2009.

\bibitem{Sun12}
Q.~Sun, ``Recovery of sparsest signals via $l_p$-minimization,'' \emph{Appl.
  Comput. Harmon. Anal.}, vol.~32, no.~3, pp. 329--341, 2012.

\bibitem{LaiL11}
M.-J. Lai and L.~Y. Liu, ``A new estimate of restricted isometry constants for
  sparse solutions,'' \emph{Online
  http://www.math.uga.edu/~mjlai/papers/LaiLiu11.pdf}, 2011.

\bibitem{Wuc13}
R.~Wu and D.-R. Chen, ``The improved bounds of restricted isometry constant for
  recovery via $l_p$-minimization,'' \emph{IEEE Trans. Inf. Theory}, vol.~59,
  no.~9, pp. 6142--6147, 2013.

\bibitem{DauDFG10}
I.~Daubechies, R.~Devore, M.~Fornasier, and S.~Gunturk, ``Iteratively
  reweighted least squares minimization for sparse recovery,'' \emph{Comm. Pure
  Appl. Math}, vol.~63, no.~1, pp. 1--38, 2010.

\bibitem{Cha07}
R.~Chartrand, ``Exact reconstruction of sparse signals via nonconvex
  minimization,'' \emph{IEEE Signal Processing Letters}, vol.~14, no.~10, pp.
  707--710, 2007.

\bibitem{BahR13}
S.~Bahmani and B.~Raj, ``A unifying analysis of projected gradient descent for
  $l_p$-constrained least squares,'' \emph{Appl. Comput. Harmon. Anal.},
  vol.~34, no.~11, pp. 366--378, 2013.

\bibitem{Fou10}
S.~Foucart, ``Sparse recovery algorithms: sufficient conditions in terms of
  restricted isometry constants,'' \emph{Approximation Theory XIII: San Antonio
  2010, Springer Proceedings in Mathematics}, vol.~13, pp. 65--77, 2010.

\end{thebibliography}
\end{document}